\documentclass[onecollarge]{svjour2}
\usepackage{epsfig}
\usepackage{amsmath}
\usepackage{amssymb}
\usepackage{graphicx}
\usepackage{color}
\usepackage{epstopdf}

\journalname{Few-Body Syst}

\begin{document}

\title{Bound Chains of Tilted Dipoles in Layered Systems}

\author{A.~G. Volosniev \and J.~R. Armstrong \and D.~V. Fedorov \and A.~S. Jensen \and N.~T. Zinner}

\institute{ Department of Physics and Astronomy,
         Aarhus University, DK-8000 Aarhus C, Denmark }

\date{\today}

\maketitle

\begin{abstract}
Ultracold polar molecules in multilayered systems have been experimentally realized very recently.
While experiments study these systems almost exclusively through their chemical reactivity, the 
outlook for creating and manipulating exotic few- and many-body physics in dipolar systems is 
fascinating. Here we concentrate on few-body states in a multilayered setup. We exploit the 
geometry of the interlayer potential to calculate the two- and three-body chains
with one molecule in each layer. The focus is on dipoles that are aligned at some angle
with respect to the layer planes by means of an external eletric field. The binding energy
and the spatial structure of the bound states are studied in several different ways using
analytical approaches. The results are compared to stochastic variational calculations and 
very good agreement is found. We conclude that approximations based on harmonic oscillator
potentials are accurate even for tilted dipoles when the geometry of the potential 
landscape is taken into account. 
\end{abstract}

\maketitle

\section{Introduction}
The prospect of extending the potential of ultracold quantum gases as a 
quantum simulation device into the realm of long-range interacting systems
has received a massive boost with the recent success in producing cold
polar molecules in their rotational and vibrational ground 
state \cite{ospelkaus2008,ni2008,deiglmayr2008,lang2008,carr2009,ni2010,ospelkaus2010,aikawa2010,danzl2010}.
Systems of this type hold great promise for exploration of non-trivial 
few- and many-body dynamics with both fermions and bosons \cite{baranov2008,lahaye2009}.
Three-dimensional samples with dipoles can unfortunately be highly unstable
to chemical reactions losses \cite{ospelkaus2010}. Trapping 
the dipoles in low-dimensional geometries have therefore been proposed
to counteract this problem \cite{fischer2006,wang2006,micheli2007,buchler2007,gorshkov2008} 
and recent experiments at JILA find large geometric effects that confirm a reduced loss in quasi-2D layers
\cite{miranda2011} and in three-dimensional optical lattices \cite{chotia2011}. The layered
geometry with dipolar molecules has attracted much attention and exotic many-body 
phases such as paired states \cite{bruun2008,cooper2009,potter2010,pikovski2010,zinner2010,baranov2011,levinsen2011},
interlayer coherence akin to ferromagnetism \cite{sarma2009}, anisotropic 
superfluidity \cite{ticknor2011}, density waves \cite{sun2010,yamaguchi2010,zinner2011,babadi2011,sieberer2011,parish2011}, 
and non-trivial Fermi liquids \cite{chan2010,kestner2010,lu2011}. However, the few-body structure is
equally intriguing and bound complexes have been found with dipoles perpendicular to the plane
in bi- \cite{shih2009,jeremy2010,klawunn2010a,fedorov2011,zinner2011c} and multilayers \cite{armstrong2011,artem2011c}, 
for tilted dipoles in single \cite{cremon2010} and bilayer setups \cite{artem2011b,artem2011a}, and in one-dimensional 
tubular geometries \cite{deu2010,santos2010,wunsch2011,zinner2011b,mekhov2011}.

\begin{figure}[htb!]
\epsfig{file=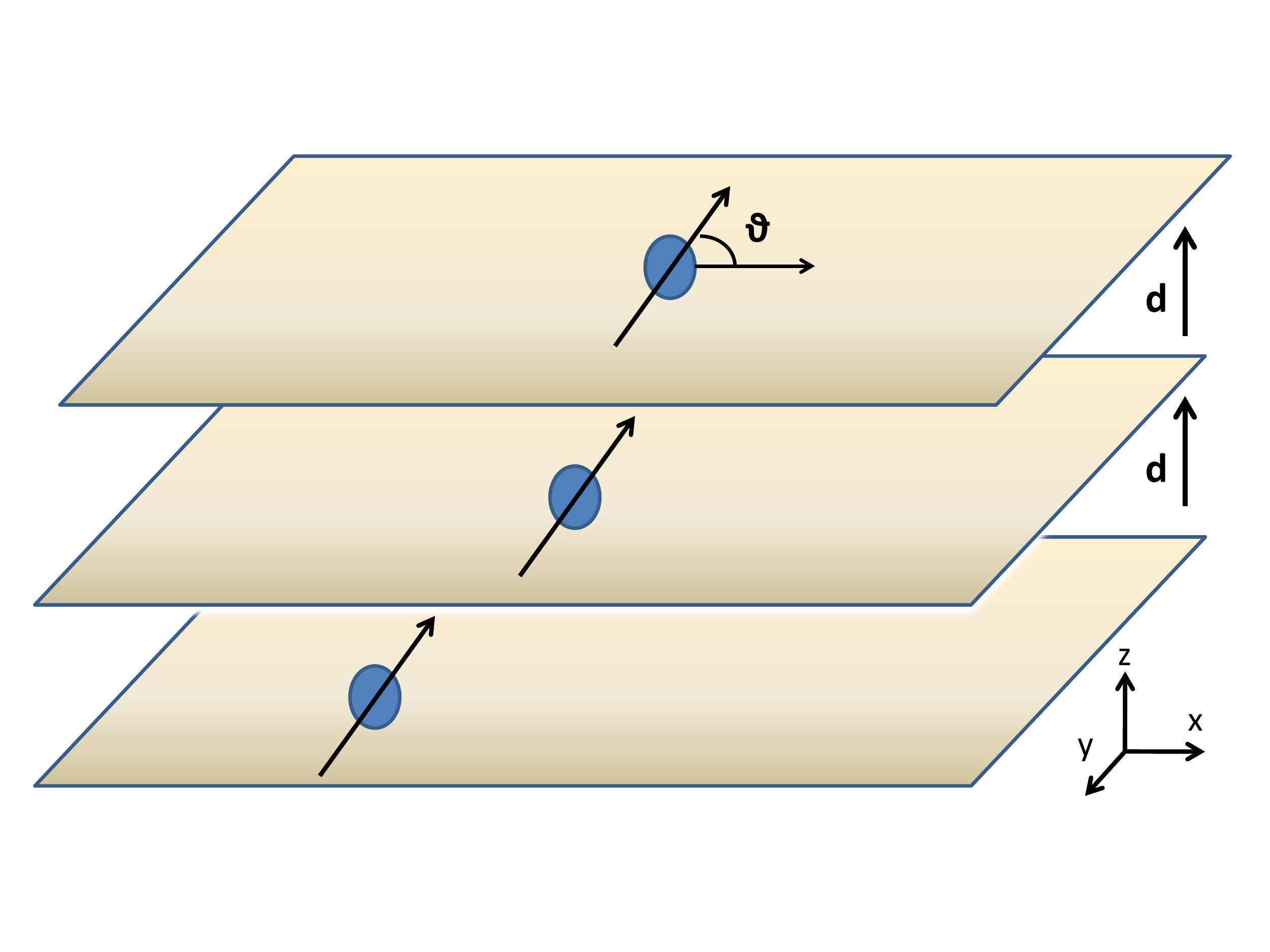,scale=0.4}
\centering
\caption{Schematic picture of the trilayer geometry. The layer distance is $d$ and the 
coordinates we use are indicated in the bottom right-hand corner. The angle between
the layer planes and the externally aligned dipole moment is $\theta$ and is taken to
lie in the $xz$-plane.}
\label{schematic}
\end{figure}

In this paper we focus on a few-body problem in a multilayered stack of planes when the 
dipoles are not perpendicular to the layer planes. Here we focus on the case of three
adjacent planes, but we will comment on the implications for more layeres at the 
end of the presentation. A schematic view of the geometry and the dipoles is
shown in Fig.~\ref{schematic}. The energy and spatial structure of these 
two- and three-body dipolar complexes is our interest and we will use analytical 
approximations to the potentials to construct a harmonic model that is 
exactly solvable. However, this requires careful consideration of the 
geometry of the potential landscape of the molecules. 
To check how well
an exact harmonic model can describe the system, we performed full stochastic 
variational calculations in a novel manner that takes the deformation of the 
potential for non-perpendicular dipoles into account. 
Our findings clearly demonstrate that the harmonic approach is very accurate
for intermediate and strong dipole strength when the potential geometry is 
included properly. This holds for most
tilting angles and only breaks down when the dipoles are oriented parallel
to the layer. Our finding imply that away from the weakly interacting limit, 
the harmonic approximation works well in layered dipolar systems also in the 
case where the dipoles are tilted. This can be used in exact $N$-body 
harmonic Hamiltonian studies \cite{nbody2011,method2012} of both energetics, 
thermodynamics, and instabilities in larger 
systems \cite{wang2006,armstrong2011,armstrong2012} with more layers
which is the case for recent experiments with around 30 layers \cite{miranda2011}.

\section{Model and results}
The dipolar potential between two dipoles with dipole moment $D$
in adjacent layers seperated by a distance $d$ is
\begin{equation}
V(x,y)=D^2\frac{x^2+y^2+d^2-3(x\cos(\theta)+d\sin(\theta))^2}{(x^2+y^2+d^2)^{5/2}},
\label{ref-1}
\end{equation}
where $\theta$ is the angle between the dipole direction and the layer plane as 
shown in Fig.~\ref{schematic}. Below we will use $d$ as the unit for length, 
and $U=mD^2/\hbar^2d$ as the dipolar strength with $m$ the mass of the molecules.
Energies will then be in units of $\hbar^2/md^2$. The molecules used in recent 
experiments \cite{miranda2011} are KRb which has a maximum dipole of 
0.566 Debye. With the setup in Ref.~\cite{miranda2011}, this gives a maximum
$U\sim 1.2$ which is in the weak-coupling limit. However, molecules like LiCs
or RbCs which are currently being cooled have much larger dipole moments \cite{dulieu2006}
and increasing $U$ by factor of 10 or more should be possible with those 
species.
Here we consider the strict two-dimensional limit
where we can neglect the width of the layers. This holds when the optical lattice
potential producing the layers is deep. Corrections to this can easily be taken
into account by integrating out the transverse ($z$-direction) degree of freedom
with the wave function of the lattice (simply a Gaussian when the lattice is deep).

\begin{figure}[htb!]
\epsfig{file=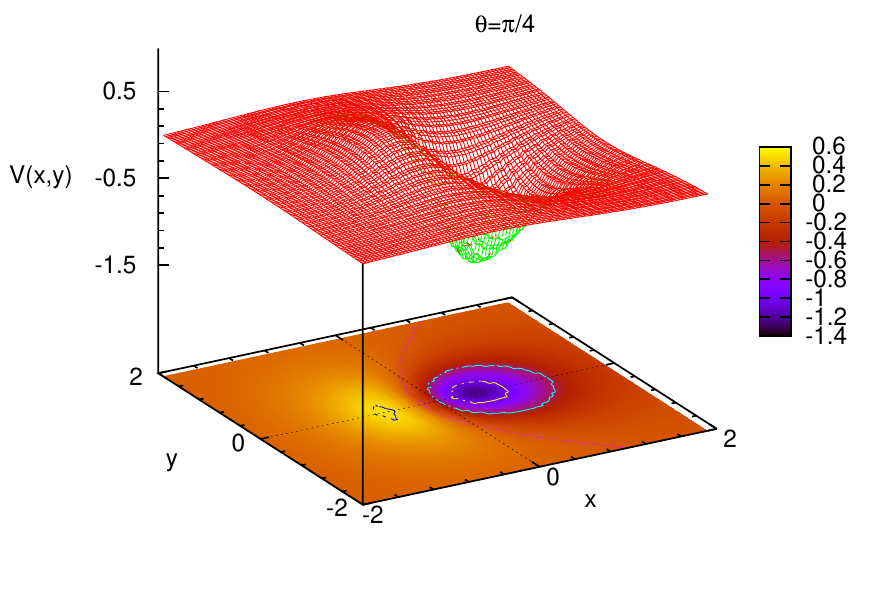,scale=1}
\centering
\caption{Contour plot of the interlayer potential for $\theta=\pi/4$ and $U=1$.
The coordinates $x$ and $y$ are measured in units of $d$. Notice the strong
asymmetry of the potential landscape in the $x$-direction. The potential has 
reflection symmetry along the $y$-direction.}
\label{3Dpot}
\end{figure}

For $\theta=\pi/2$ the potential is spherically symmetric, but for all other
angles there is a distinct geometric structure. In Fig.~\ref{3Dpot} we show a 
contour plot of the potential for $\theta=\pi/4$ where a clear asymmetric 
structure in the $x$-direction is visible. It is this fact that we can exploit
to build an accurate analytical approximation to the ground-state wave function
of the two- and three-body systems. Notice that there will be a three-body
bound state for {\it any} value of $U$ in this setup with one molecule in 
each of the three layers. This can be understood from the fact that the 
two-body bound state in the case of two layers always exists as discussed 
in Refs.~\cite{artem2011a,artem2011b}. Adding the third layer and molecule
will then render the binding energy even smaller, i.e. $E_3<E_2$. This 
is a simple generalization of the proof given in Ref.~\cite{cou1983}
to the layered dipolar case where $\int dxdy V(x,y)=0$.

\begin{figure}[htb!]
\epsfig{file=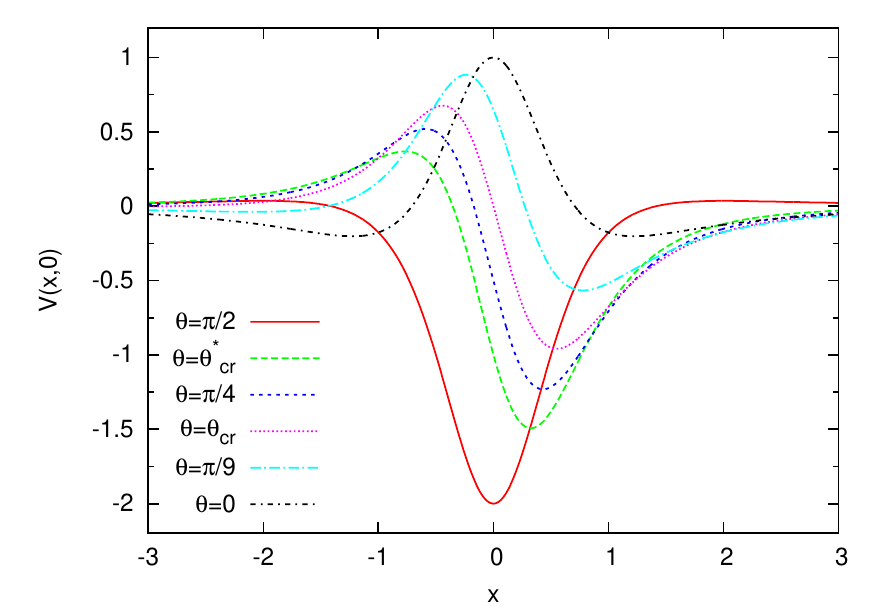,scale=1}
\centering
\caption{Interlayer potential for $y=0$ and for different values of $\theta$ for $U=1$. The two
special angles are defined through the relations $\cos^2\theta_{c}^{*}=\tfrac{1}{3}$ and
$\sin^2\theta_{c}^{}=\tfrac{1}{3}$.}
\label{xpot}
\end{figure}

To further illustrate the interesting geometrical structure of the potential 
along the $x$-direction, Fig.~\ref{xpot} displays the potential for $y=0$ 
and various values of $\theta$ between zero and $\pi/2$. For $\theta=0$, 
there are two global minima located on each side of $x=0$, whereas for 
$\theta=\pi/2$ we have a single minimum at $x=0$ and also cylindrical 
symmetry. In the regime $0< \theta\leq \theta_{c}^{*}\sim 0.955$ 
($\cos^2\theta_{c}^{*}=\tfrac{1}{3}$), 
the potential has two minima, one is located at $x>0$ and is deep, while
the other one has a small minimum for $x<0$. At the intermediate angle $\theta_{c}$ 
($\sin^2\theta_c=\tfrac{1}{3}$) the potential is maximally asymmetric 
since here the monopole term vanishes \cite{artem2011b}. The monopole 
term can be deduced by expanding the potential in a two-dimensional 
spherical expansion via terms of the form $\cos\phi$ where $x=r\cos\phi$
and $y=r\sin\phi$ with $r=\sqrt{x^2+y^2}$
(the $y\leftrightarrow -y$ symmetry of the potential 
excludes $\sin\phi$ terms). The potential can be expanded using the 
basis function $1$, $\cos\phi$, and $\cos2\phi$ only, these correspond
to a monopole, dipole, and quadrupole term as discussed in 
detail in Ref.~\cite{artem2011b}.

We want to exploit the geometry of the interlayer potential to determine the energy
and structure of the bound states. This can be done by considering expansions of the 
Hamiltonian that use the minima of the potential (on the $x>0$ side) as the starting
point. Denote the minimum position by $(x,y)=(a,0)$, where $a$ is defined through the 
condition $\tfrac{\partial V(x,0)}{\partial x}=0$. The equation determining $a$ is
\begin{equation}
15a(a\cos(\theta)+d\sin(\theta))^2-3a(a^2+d^2)-6\cos(\theta)(a^2+d^2)(a\cos(\theta)+d\sin(\theta))=0,
\label{ref-2}
\end{equation}
from which one can easily see that $a$ scales as $d$, e.g. $a=d a_0$, where $a_0$ is the solution of Eq.~\eqref{ref-2} for $d=1$.

The potential can now be written in term of the variable $w=x-a$ centered at the minimum 
\begin{equation}
V(w,y)=D^2\frac{w^2+y^2+d^2+a^2+2wa-3(w\cos(\theta)+d\sin(\theta)+a\cos(\theta))^2}{(w^2+y^2+d^2+a^2+2wa)^{5/2}}.
\label{ref-3}
\end{equation}
If we expand this to second order in $w$ and $y$ we have
\begin{equation}
V(w,y)=v(d)+\alpha(d)w^2+\beta(d) y^2.
\label{ref-4}
\end{equation} 
where 
\begin{eqnarray}
&v(d)=\frac{D^2}{d^3}v_0\;,v_0=\frac{a_0^2+1-3(\sin(\theta)+a_0\cos(\theta))^2}{(a_0^2+1)^{5/2}},&\label{ref-5}\\
&\alpha(d)=\frac{D^2}{d^5}\alpha_0,&\label{ref-6}\\
&\alpha_0=
\frac{1-3\cos^2(\theta)+\frac{(30a_0^2-5)(1+a_0^2-3(\sin(\theta)+
a_0\cos(\theta))^2)}{2(a_0^2+1)}-
5\frac{a_0(2a_0-6\cos(\theta)(a_0\cos(\theta)+\sin(\theta)))}{2(a_0^2+1)^2}}{(a_0^2+1)^{5/2}},&\nonumber\\
&\beta(d)=\frac{D^2}{d^5}\beta_0\;,
\beta_0=-\frac{3}{2}\frac{1+a_0^2-5(a_0\cos(\theta)+\sin(\theta))^2}{(a_0^2+1)^{7/2}}
\label{ref-7}
\end{eqnarray} 
Here we have used the scaling property $a=da_0$ to extract the dependence on $d$ in each term. This will be convenient below.

\subsection{Harmonic three-body chains}
\begin{figure}
\centering
\epsfig{file=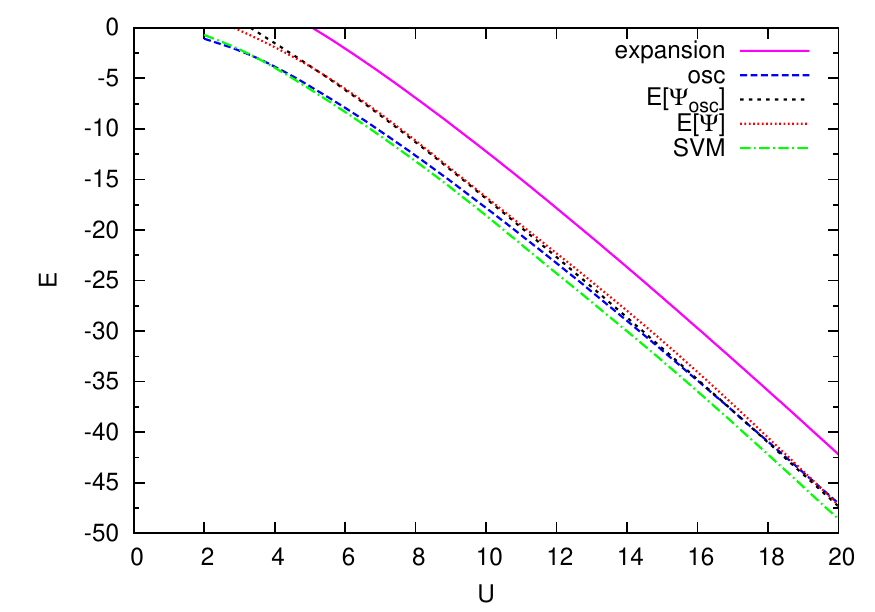,scale=1}
\caption{Bound state energy of the three-body chain as function of dipolar strength $U$ for $\theta=\tfrac{\pi}{2}$.
The dot-dashed line (SVM) is the value obtained from a numerical stochastic variational calculation.
The full line (expansion) shows the result of Eq.~\eqref{ref-11}, while the dotted ($E[\Psi]$) shows the result
of using the wave function of Eq.~\eqref{ref-12} as a variational guess for the real dipolar potential of Eq.~\eqref{ref-1}.
The long-dashed (osc) is the optimized oscillator 
based on Eq.~\eqref{ref-17}, while the short-dashed ($E[\Psi_\textrm{osc}]$) uses the wave function in Eq.~\eqref{ref-16}
as a variational guess.}
\label{pi2fig}
\end{figure}

\begin{figure}
\centering
\epsfig{file=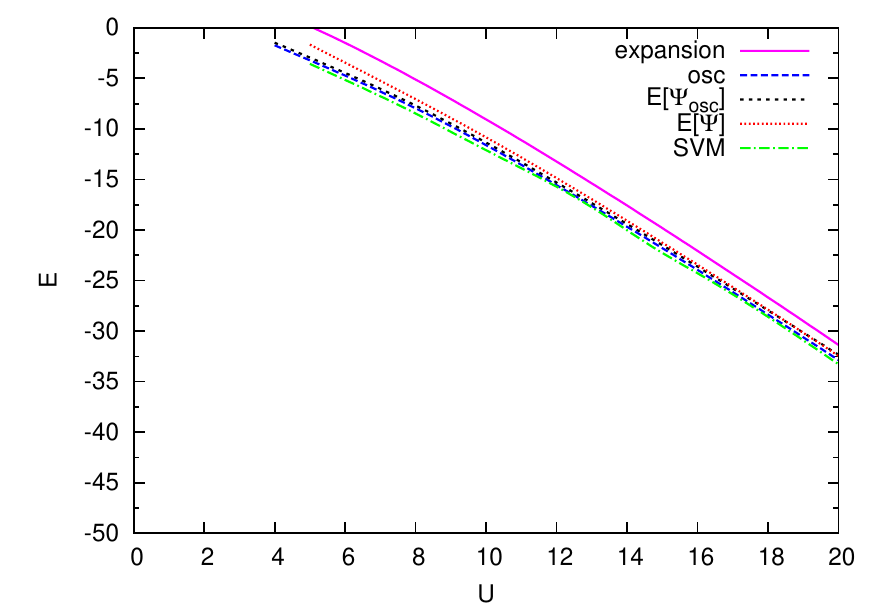,scale=1}
\caption{Same as Fig.~\ref{pi2fig} but for $\theta=\theta_{c}^{*}$.}
\label{critfig}
\end{figure}

Consider three adjacent layers with interlayer distance $d$ as shown in Fig.~\ref{schematic}. The potential for 
a three-particle chain system with one particle in each layer using the harmonic approximation above is
\begin{eqnarray}
&V(x_1,x_2,x_3,y_1,y_2,y_3)=\frac{17D^2v_0}{8d^3}+\frac{D^2\alpha_0}{d^5}(x_1-x_2-da_0)^2+\frac{D^2\beta_0}{d^5}(y_1-y_2)^2&\nonumber\\
&+\frac{D^2\alpha_0}{d^5}(x_2-x_3-da_0)^2+\frac{D^2\beta_0}{d^5}(y_2-y_3)^2+\frac{D^2\alpha_0}{(2d)^5}(x_1-x_3-2 da_0)^2+\frac{D^2\beta_0}{(2d)^5}(y_2-y_3)^2.&
\label{ref-8}
\end{eqnarray} 
Notice the subtraction by $2a_0$ in the term relating the $x_1$ and $x_3$ coordinates. This reflects the expected geometrical
structure of the bound chain. Focus now on the $x$-dependent terms. Introducing the relative coordinate set $x_1-x_3=\sqrt{2}q_{1x}$, 
$ x_1-x_2=\sqrt{\frac{1}{2}}q_{1x}+\sqrt{\frac{3}{2}}q_{2x}$, and $x_3-x_2=-\sqrt{\frac{1}{2}}q_{1x}+\sqrt{\frac{3}{2}}q_{x2}$,
we have 
\begin{eqnarray}
&V_x(q_{1x},q_{2x})=\frac{D^2\alpha_0}{d^5}(\sqrt{\frac{1}{2}}q_{1x}+\sqrt{\frac{3}{2}}q_{2x}-da_0)^2
+\frac{D^2\alpha_0}{d^5}(-\sqrt{\frac{1}{2}}q_{1x}+\sqrt{\frac{3}{2}}q_{2x}+da_0)^2&\nonumber\\&+\frac{D^2\alpha_0}{(2d)^5}(\sqrt{2}q_1-2 da_0)^2
=\frac{17}{8}\frac{D^2\alpha_0}{d^5} (da_0)^2+\frac{17}{16}\frac{D^2\alpha_0}{d^5} q_{1x}^{2}+3\frac{D^2\alpha_0}{d^5} q_{2x}^{2}-\frac{17}{8}\frac{D^2\alpha_0}{d^5}\sqrt{2}q_{1x}da_0&\nonumber\\
&=\frac{17}{16}\frac{D^2\alpha_0}{d^5}(q_{1x}-\sqrt{2}da_0)^2+3\frac{D^2\alpha_0}{d^5}q_{2x}^{2}.&
\label{ref-10}
\end{eqnarray} 
The ground-state energy in this oscillator potential is simply
$E_x=\frac{\hbar^2}{md^2}\sqrt{U\alpha_0}\left(\sqrt{\frac{3}{2}}+\sqrt{\frac{17}{32}}\right)$
with one term from each of the one-dimensional oscillator coordinates $q_{1x}$ and $q_{2x}$. 
The $y$-direction will give an identical contribution with $\beta_0$ instead of $\alpha_0$ and 
the final result for the ground-state energy of the three-chain, $E$, within this harmonic approximation is
\begin{equation}
\frac{md^2}{\hbar^2}E=\left(\sqrt{\frac{3}{2}}+\sqrt{\frac{17}{32}}\right)\,\left(\sqrt{\alpha_0}+\sqrt{\beta_0}\right)\,\sqrt{U}
+\frac{17}{8}v_0\,U.
\label{ref-11}
\end{equation}
Below we will refer to this result for the energy as 'expansion' since it corresponds to a naive expansion 
of the potential around its minimum position to second order. The energies obtained from 
Eq.~\eqref{ref-11} are shown in Figs.~\ref{pi2fig} and \ref{critfig}. For $\theta=\theta_{c}^{*}$, we 
find $a_0=\tfrac{3\sqrt{17}-5}{2^{9/2}}$ which gives $\alpha_0\sim 2.66$, $\beta_0\sim 4.11$, and $v_0\sim -1.47$.
In the perpendicular case where $\theta=\pi/2$, we have $a_0=0$, $\alpha_0=\beta_0=6$, and $v_0=-2$.

The corresponding wave function in the 'expansion' method can be written
\begin{eqnarray}
&\Psi(q_{1x},q_{2x},q_{1y},q_{2y})=N\exp\left[-\frac{1}{2}\sqrt{6U\alpha_0}\left(\frac{q_{2x}}{d}\right)^2
-\frac{1}{2}\sqrt{\frac{17U\alpha_0}{8}}\left(\frac{q_{1x}}{d}-\sqrt{2} a_0\right)^2\right.&\nonumber\\
&\left.-\frac{1}{2}\sqrt{6U\beta_0}\left(\frac{q_{2y}}{d}\right)^2
-\frac{1}{2}\sqrt{\frac{17U\beta_0}{8}}\left(\frac{q_{1y}}{d}\right)^2\right],&
\label{ref-12}
\end{eqnarray} 
where the normalization constant is $N^{2}=\sqrt{\tfrac{51}{4}}\sqrt{\alpha_0\beta_0}U/\pi^2$. This can 
be used to improve the approximation by considering instead the expectation value of the full 
potential from Eq.~\eqref{ref-1} and calculating $E[\Psi]=\frac{<\Psi|H|\Psi>}{<\Psi|\Psi>}$. 
The result of this proceduce is also shown on Figs.~\ref{pi2fig} and \ref{critfig}. In comparison
to the result from Eq.~\eqref{ref-11}, $E[\Psi]$ is closer to the full numerical result 
based on the stochastic variational method (SVM) \cite{artem2011b}. This implies that a harmonic
oscillator wave function is a good variational guess.
The difference between Eq.~\eqref{ref-11} and $E[\Psi]$ arises mainly
because the kinetic energy is overestimated in the naive expansion.

To investigate further the performance of harmonic approximations to the dipolar potential, we now 
consider an optimization on the two-body interlayer potential where the minimum position $a$ is a
variational parameter. We consider the energy functional 
\begin{equation}
E[G]=\frac{<G|H|G>}{<G|G>},
\label{ref-15}
\end{equation}
where $H$ is the full Hamiltonian with the real potential from Eq.~\eqref{ref-1} and 
\begin{equation}
G=\exp\left(-\frac{\tilde\alpha}{2}(\frac{x_1-x_2-\tilde a}{d})^2-\frac{\tilde\beta}{2} (\frac{y_1-y_2}{d})^2\right),
\label{ref-16}
\end{equation}
is a two-body wavefunction. Fixing $U$ and $\theta$, we can now minimize $E[G]$ to find the optimal 
set of $\tilde\alpha$, $\tilde\beta$, and $\tilde a$. This suggests what the optimal oscillator potential should 
be by replacing the real dipolar potential by a harmonic one of the form
\begin{equation}
V_{osc}=\frac{\hbar^2}{2md^2}\tilde\alpha^2(\frac{x_1-x_2-\tilde a}{d})^2+\frac{\hbar^2}{2md^2}\tilde\beta^2(\frac{y_1-y_2}{d})^2,
\label{ref-17}
\end{equation}
which will give the energy $E_\textrm{osc}=\tfrac{1}{2}\tfrac{\hbar^2}{md^2}\left(\tilde\alpha+\tilde\beta\right)$. 
This optimized harmonic potential for two particles in two adjacent layers can now be used as the
starting point for the three-body calculations by replacing $\alpha_0$, $\beta_0$, 
and $a_0$ in Eq.~\eqref{ref-8} by $\tilde\alpha$, $\tilde\beta$, and $\tilde a$ (and dropping the constant term $v_0$).
The results of this calculation is shown on Figs.~\ref{pi2fig} and \ref{critfig} where it is 
labeled 'osc'. Just like above, one can use the three-body wave function obtained from this optimized 
choice of harmonic oscillator to calculate the expectation value of the real dipolar potential, 
the results of which is labeled '$E[\Psi_\textrm{osc}]$'.

From Figs.~\ref{pi2fig} and Figs.~\ref{critfig} we see that our harmonic approximations are
close (expect for the naive expansion) for $U\gtrsim 2$ ($\theta=\pi/2$) and $U\gtrsim 4$ 
($\theta=\theta_{c}^{*}$). Below these values, the wave function spread out too much to be 
described reliably by a single gaussian. Clearly, the naive expansion approach result 
from Eq.~\eqref{ref-11} performs the worst, although it appears to do somewhat better for
tilted dipoles ($\theta<\pi/2$). However, using the wave function from Eq.~\eqref{ref-12}
as a variational guess is a great improvement as demonstrated by the $E[\Psi]$ results.
The optimzed oscillator results 'osc' and the corresponding variational guess based
on this wave function, $E[\Psi_\textrm{osc}]$, are both better than the expansion, 
and they differ only slightly. This is to be expected since the optimized oscillator
will take as much of the real potential into account as possible using the 
three (essentially free) parameters $\tilde\alpha$, $\tilde\beta$, and $\tilde a$.

The fact that the result of Eq.~\eqref{ref-11} is rather poor, yet the wave function
Eq.~\eqref{ref-12} is a good variational guess, is interesting. Note that the 
slope of the exact SVM and Eq.~\eqref{ref-11} is almost the same. This demonstrates
that one can get a very accurate approximation for the energy
by merely shifting the expansion result. This strategy has been pursued for 
$\theta=\pi/2$ recently to study the properties of chains in multiple layers \cite{armstrong2011,armstrong2012}.
The current results show that this approximation can be extended to the non-perpendicular 
case also.

\begin{figure}
\centering
\epsfig{file=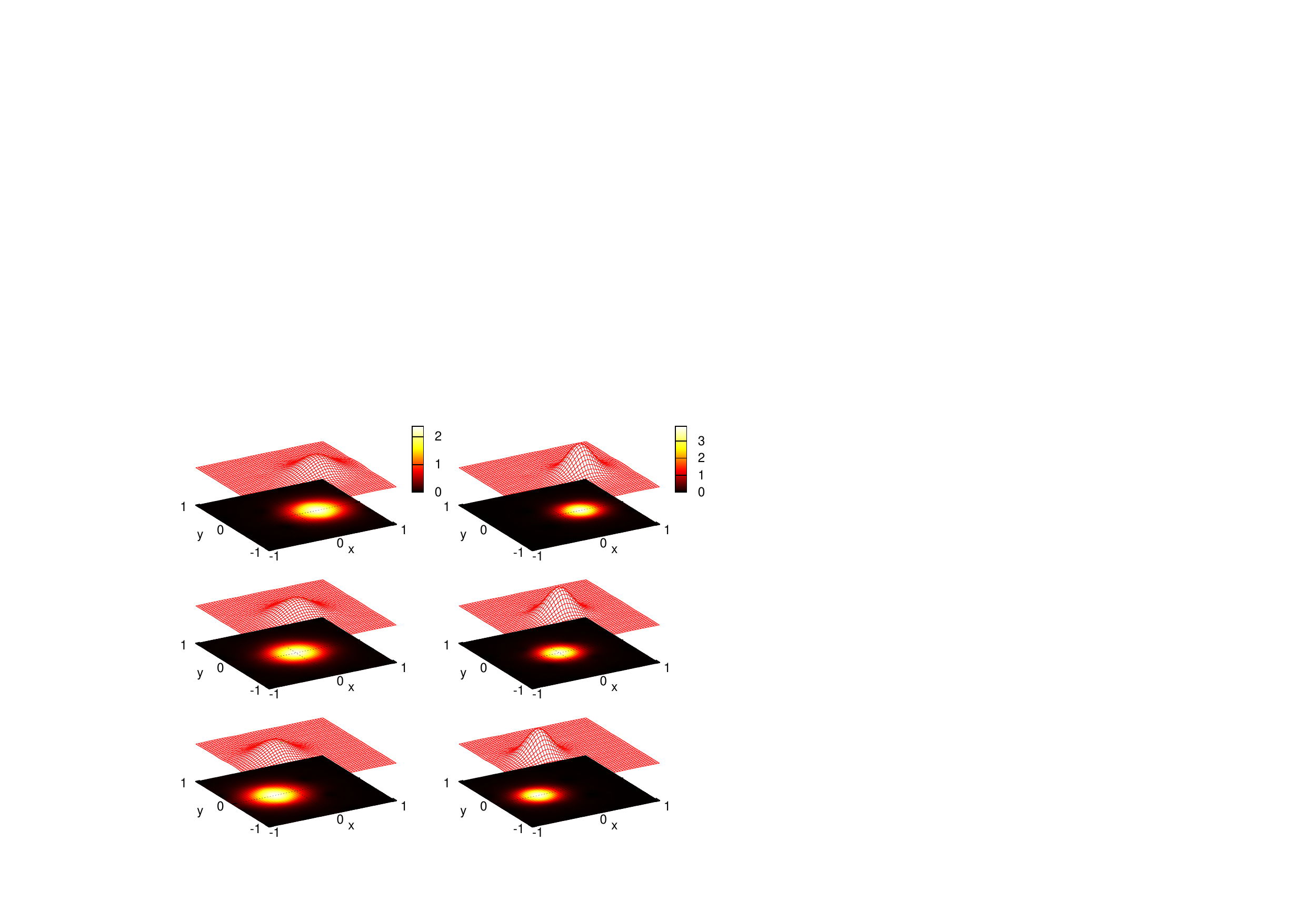,scale=1}
\caption{Probability densities for the molecules in each of the 
layeres when $\theta=\theta_{c}^{*}$. Left side has $U=5$, while the right side
has $U=15$.}
\label{wavefig}
\end{figure}

To further investigate the structure we now look at the probability distribution 
for the three-body chain. The (partial) probability distribution for the first
molecules is given by
\begin{equation}
F(x_1,y_1)=\int \vert\Psi(q_{1x},q_{2x},q_{1y},q_{2y})\vert^2 
\delta(\frac{\vec r_1+\vec r_2+\vec r_3}{\sqrt{3}}) \mathrm{d}\vec r_1 \mathrm{d}\vec r_2\mathrm{d}\vec r_3,
\end{equation}
where we use the delta function to eliminate the integration over the coodinates of molecule 1, $\vec r_1$, 
and similarly for the other molecules. In Fig.~\ref{wavefig} we show the probability distributions
in each layer for $\theta=\theta_{c}^{*}$ with $U=5$ and $U=15$ based on the gaussian
wave functions used above (expansion method). This wave function very accurately describes the 
spatial structure of the full solution. The figure clearly demonstrates the 
geometry of the dipolar potential with maximum in the middle of the center layer and 
displaced maxima in the two outer layers. This trend will continue for more
than three layers. Notice also how the extend of the probability shrinks
as $U$ increases in response to the increased localization in the potential 
minimum.

While we have only considered $\theta=\pi/2$ and $\theta=\theta_{c}^{*}$ explicitly, similar
results are obtained for a large range of angles $\theta>0$. Only close to $\theta=0$ 
do we run into problems since the potential has a two-minima structure which is not
well described with a single gaussian for each molecule. Here one should presumably use at 
least two gaussians, one for each minimum. This is a subject for future investigations.

\section{Discussion}
We have studied polar molecules in a geometry consisting of equally spaced two-dimensional 
planes which is predicted to have a rich collection of few-body states due to the 
long-range and anisotropic dipole-dipole interaction between the molecules. In particular,
the long-range character reaches across the different layers and provides an attractive force
that can hold non-trivial bound states of several molecules. Here we have considered a simple
case where there are three layeres and one molecule in each layer. This means that there are 
only interlayer interactions to worry about. When the dipole moment of the molecules
are aligned perpendicular to the layers, a three-body bound chain is always present. 
While this result is expected to hold also when the dipoles are tilted away from 
perpendicular, no systematics has been considered thus far. Here we extend these
results and show that for tilted dipoles the three-body chain persists and it 
can be well approximated by various gaussian wave function schemes. This 
conclusion can be easily extended to chains in setups with more than three layers.

Our results demonstrate that harmonic approximations to the dipolar interaction
are very accurate in determining the bound state energy and also the 
spatial structure of the wave function in the intermediate and strong 
dipolar interaction regime also when the dipole moment is tilted with 
respect to the layers. Perhaps somewhat surprisingly, the harmonic
approximation seems to work better around the critical angle (defined 
as the angle where two dipoles on a line will have zero interaction)
than in the perpendicular case. We thus conclude that harmonic
approximations are a viable and accurate strategy when studying the 
rich few-body structure of polar molecules in low-dimensional and/or
lattice geometries.

\end{document}